\documentclass{iopart}
\usepackage{graphicx}
\usepackage{latexsym}
\usepackage{iopams, amsopn, amstext}
\usepackage{colonequals}
\usepackage{dcolumn}
\usepackage{setspace}
\usepackage{subfigure}
\usepackage{bm}
\usepackage{url}

\graphicspath{{Figures/}}

\def\bh#1{black hole#1
  (BH#1)\gdef\bh{BH}}
\def\bbh#1{binary black hole#1
  (BBH#1)\gdef\bbh{BBH}}
\def\gw#1{gravitational wave#1
  (GW#1)\gdef\gw{GW}}
  \def\qnm#1{quasi-normal mode#1
  (QNM#1)\gdef\qnm{QNM}}
\def\gr#1{general relativity#1
  (GR#1)\gdef\gr{GR}}
   
\begin{document}

\title{Decoding the final state in binary black hole mergers}
\author{James Healy$^{1,2}$, Pablo Laguna$^{1}$ and Deirdre Shoemaker$^{1}$}

\address{$^1$Center for Relativistic Astrophysics and
School of Physics, 
Georgia Institute of Technology, Atlanta, GA 30332 \\
$^2$ Center for Computational Relativity and Gravitation,
School of Mathematical Sciences,
Rochester Institute of Technology, 85 Lomb Memorial Drive, Rochester,
 New York 14623}

\begin{abstract}
We demonstrate that in binary black hole mergers there is a direct correlation between the frequency of the gravitational wave at peak amplitude and the mass and spin of the final black hole. This correlation could potentially assist with the analysis of gravitational wave observations from binary black hole mergers.
\end{abstract}

\emph{Introduction:} Because of their expected luminosity, mergers of \bbh{s} will be main targets of  \gw{}  interferometric detectors like LIGO~\cite{0264-9381-27-8-084006}, Virgo~\cite{0264-9381-28-11-114002} and KAGRA~\cite{PhysRevD.88.043007}. Detecting merging \bbh{s} may not necessarily require exquisite knowledge of the \gw{} signal. Characterizing the binary (i.e. eccentricity, masses, spins, sky location, orientation and distance) is a different story. Matched filtering is currently considered our best option, but not surprisingly, the challenge in this case is constructing waveforms or templates that effectively cover the parameter space, a task very likely necessitating large amounts of computing resources.

Alternatives that reduce the computational cost in \gw{} data analysis are very desirable, in particular options that exploit bulk features or partial information about the \bbh{.} In this paper, we identify a feature in the \gw{s} emitted during the merger of  \bbh{s} that could potentially help pinpointing the mass and spin of the final \bh{}. The feature discovered is a correlation between the \qnm{} frequency $\omega_{\rm qn}$ and decay time $\tau_{\rm qn}$ (or equivalently the quality factor $Q \equiv \omega_{\rm qn}\,\tau_{\rm qn}/2$) of the final \bh{ } with the frequency $\omega_{\rm mx}$ of the \gw{} at peak amplitude. Since $\omega_{\rm qn}$ and $\tau_{\rm qn}$ (or $Q$) are related to the mass $M_{\rm h}$ and spin parameter $a$ of the final \bh{}~\cite{4690}, the correlation we have found also provides a connection of $\omega_{\rm mx}$ with $M_{\rm h}$ and $a$. 

In retrospect, it should not be too surprising that such a correlation exists. By the time the amplitude of the \gw{} peaks, the horizons of the coalescing \bh{s} have already merged, and most of the  \emph{hair} in the binary (masses, spins, eccentricity, etc) has been lost. Nonetheless, it is interesting that, although at peak amplitude the \bh{} system is still dynamically non-linear, the frequency of the \gw{s}  is already correlated with the \qnm{} ringing of the final \bh{.}

\emph{Numerical Relativity Simulation Bank:} For over six years, our numerical relativity effort has produced an extensive archive of waveforms, with over 512 high resolution simulations of inspiraling \bbh{s}. The simulations have been obtained with our  \texttt{Maya} code~\cite{Haas:2012bk,Healy:2011ef,Bode:2011xz,Bode:2011tq,Bode:2009mt,Healy:2009zm}.
\texttt{Maya} uses the \texttt{Einstein Toolkit} \cite{et-web},
which is based on the \texttt{CACTUS} \cite{cactus-web} infrastructure
and \texttt{CARPET} \cite{Schnetter-etal-03b} mesh refinement, with thorns generated by the
\texttt{Kranc} \cite{Husa:2004ip} library.

For the present work, we considered 269 simulations, 97 with 
non-precessing and 172 with precessing binaries. 
Among the non-precessing binaries, 43 have non-spinning \bh{s}
with mass ratios ranging between 1:1 and 1:10; 39 have total \bh{} spin
aligned with the angular momentum and mass ratios ranging 
between 1:1 and 1:7; and the last 15 have total \bh{} spin anti-aligned with 
orbital angular momentum and mass ratios between 1:1 and 1:4.  
Of the non-precessing binaries, 18 have a mass ratio of 1:1.  
Regarding the 172 precessing binary simulations, individual 
\bh{} dimensionless spin parameters range between 0 and 0.7 and mass ratios 
primarily between 2:3 and 1:4, with a few equal mass and one each of 1:6 and 1:7.

For the present work, the output from the simulations we use are $M_{\rm h}$ and $a$, calculated from the apparent horizon of the final \bh{,} and the Weyl Scalar $\Psi_4$. In data analysis, it is conventional to work with the strains polarizations $h_+$ and $h_\times$, which are related to $\Psi_4$ by  $\Psi_4 = \ddot h_+ - i\,\ddot h_\times \equiv \ddot h^\star$, with star denoting complex conjugation and over-dots time derivatives. The correlation we have found shows in both $\Psi_4$ and the strains $h_+$ and $h_\times$. To avoid the inaccuracies introduced while constructing the strains, we will center the discussion around $\Psi_4$.

As customary, we decompose $\Psi_4$ into spin-weighted spherical harmonics, namely
\begin{equation}
\label{eq:Psi4Decom}
r\,M\,\Psi_4(t;\theta,\phi) = \sum_{l,m}A_{\ell m}(t)e^{i\phi_{\ell m}(t)}\, {}_{-2} Y_{\ell m}(\theta,\phi) \,,
\end{equation}
with both $A_{\ell m}$ and $\phi_{\ell m}$ real functions. In Eq.~(\ref{eq:Psi4Decom}), $M$ denotes the total mass of the \bbh{} and $r$ the distance to the binary. The frequency of the $( \ell,m )$ mode is given by $\omega_{\ell m} = \dot \phi_{\ell m}$.
To simplify notation, unless explicitly stated, we will drop the mode labels $(\ell,m)$.

Typically, the angles $\theta$ and $\phi$ in Eq.~(\ref{eq:Psi4Decom}) are relative to a coordinate system with origin at the center-of-mass of the binary and with the $z$-axis aligned with the orbital angular momentum at the beginning of the simulation. However, given our interest in the period between coalescence and  \qnm{} ringing, for the decomposition in Eq.~(\ref{eq:Psi4Decom}), we align the $z$-axis of the coordinate system with the spin vector of the final \bh{.}

\begin{figure}[tb]
\centering
\vbox{
    \includegraphics[angle=270,width=.68\linewidth]{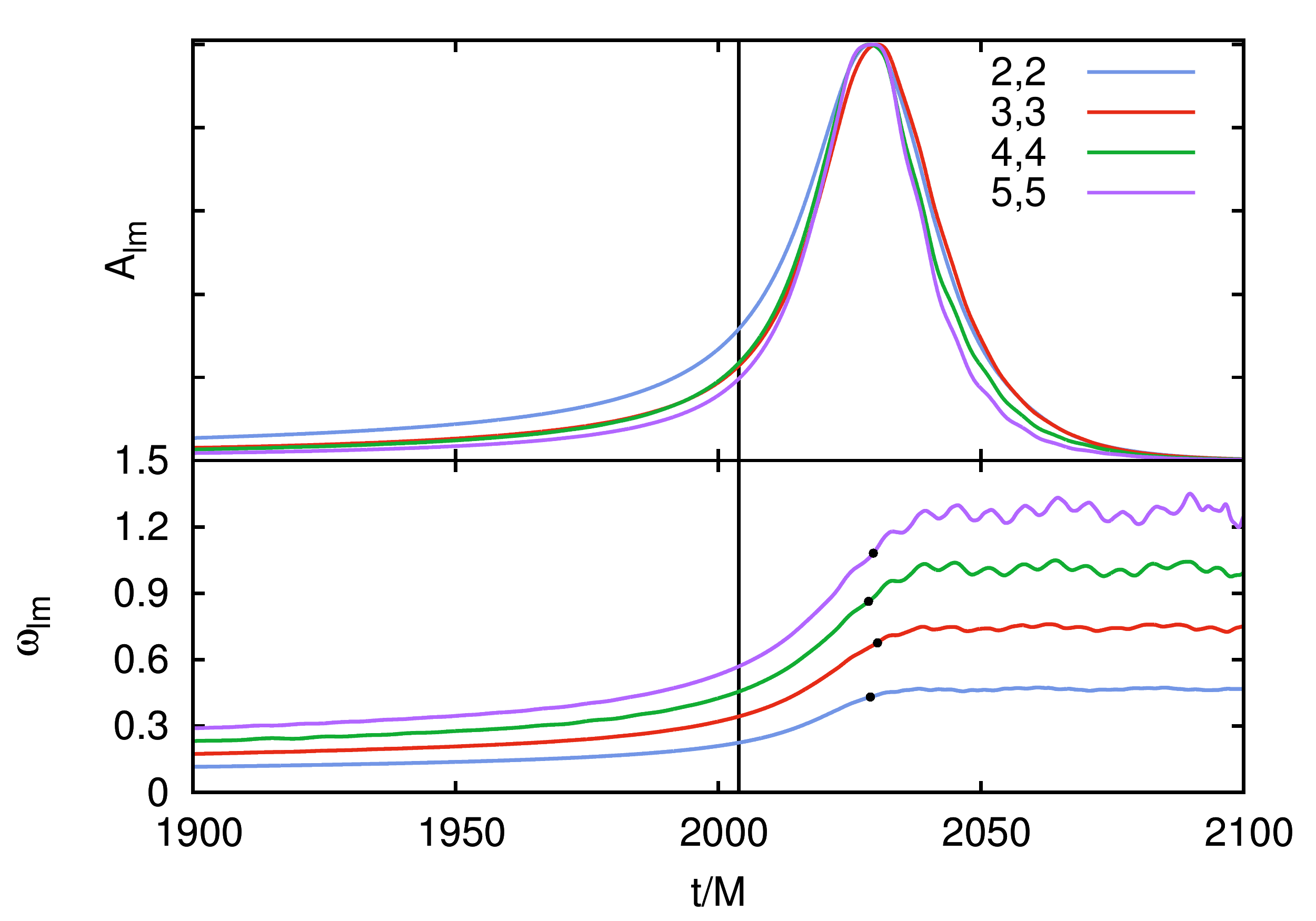}
}
\caption{Evolution of the $\Psi_4$ amplitude $A_{\ell m}$ (top panel) and frequency $\omega_{\ell m}$ (bottom panel) from the merger of a \bbh{} with equal mass, non-spinning holes. The vertical line at $t \simeq 2004\,M$ denotes the merger of the holes. The dots in the frequency lines denote the frequency $\omega_{\rm mx}$ when the corresponding amplitudes reach their maximum.  The amplitudes of the modes have been rescaled, so each of them have the same maximum values. }
\label{fig:typical}
\end{figure}

Fig.~\ref{fig:typical} shows a typical evolution of the amplitude $A_{\ell m}$ (top panel) and frequency $\omega_{\ell m}$ (bottom panel) for different $\Psi_4$ modes. The amplitudes of the modes reach a maximum at $t \sim 2030\,M$. The amplitudes have been rescaled, so each of them have the same maximum values; that is, $A_{22} = 1.48\,A_{33} = 2.32\,A_{44} = 3.21\,A_{55}$. The case depicted is that of a binary with mass ratio 1:4 and non-spinning \bh{s}. The vertical line at $t \simeq 2004\,M$ denotes the merger of the holes, i.e. when the common apparent horizon is first found. Notice that merger occurs $t \sim 25 M$ before the amplitudes reach their maximum value.
Before merger, the modes have the characteristic \emph{chirp}-like behavior (i.e. a monotonic increase of amplitude and frequency). A few $M$s after peak amplitude, the waveforms become a sum of \qnm{s} with the fundamental mode dominating, that is, $ \Psi_4 \propto e^{-t/\tau_{\rm qn}}\sin{(\omega_{\rm qn}\,t)}$. 

\begin{figure}[tb]
\centering
\vbox{
    \includegraphics[angle=270,width=.68\linewidth]{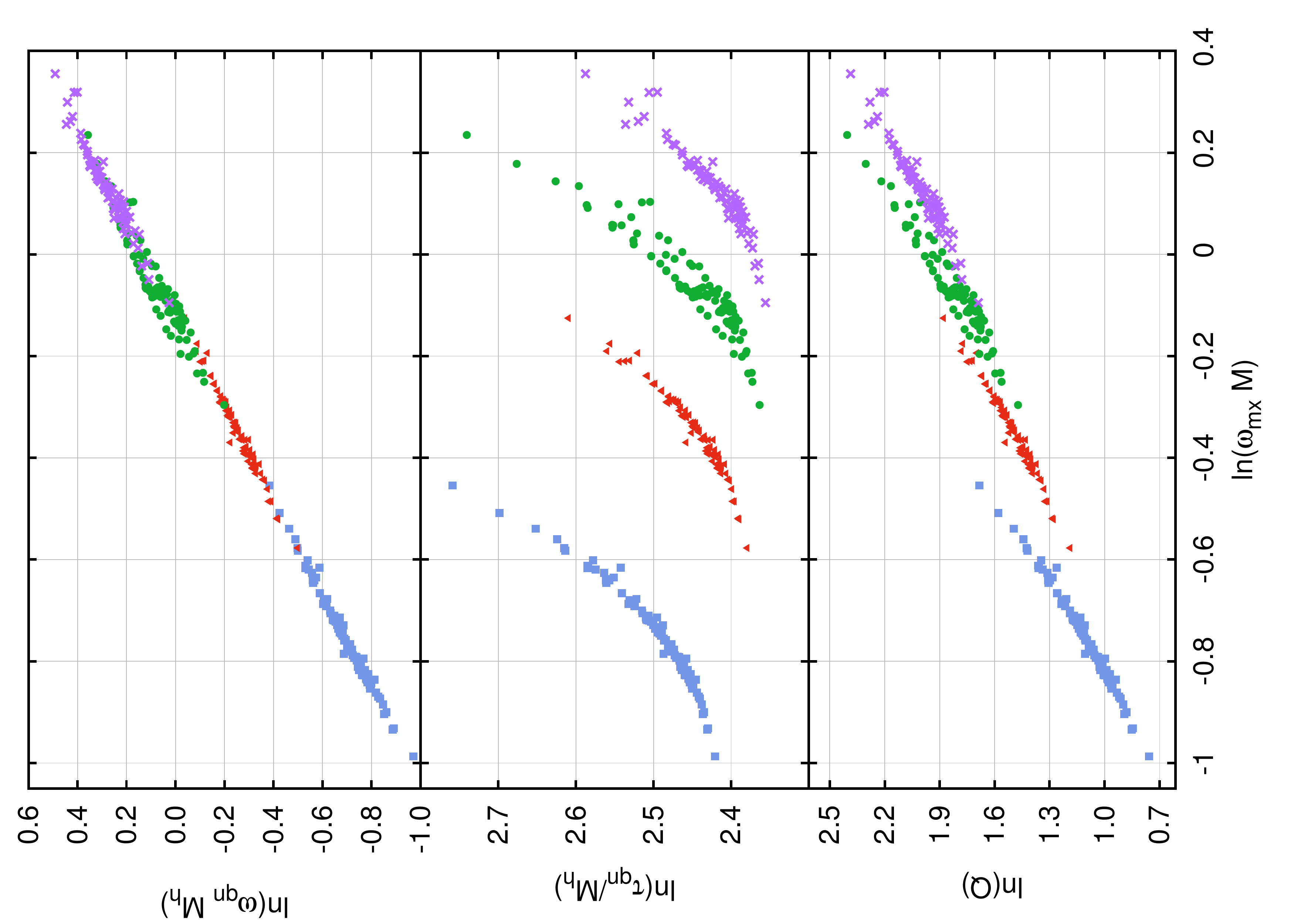}
}
\caption{The quasi-normal mode ringdown frequency 
$\hat\omega_{\rm qn}$ (top), the decay time $\hat\tau_{\rm qn}$ 
(middle), and the quality factor $Q$ (bottom) versus 
$\hat\omega_{\rm mx}$ from $\Psi_4$ for the the (2,2) [blue squares],
(3,3) [red triangles], (4,4) [green circles], and (5,5) [purple crosses] 
modes. The data only include simulations from non-precessing binaries. }
\label{fig:QNMVommax}
\end{figure}

\begin{figure}[tb]
\centering
\vbox{
    \includegraphics[angle=270,width=.68\linewidth]{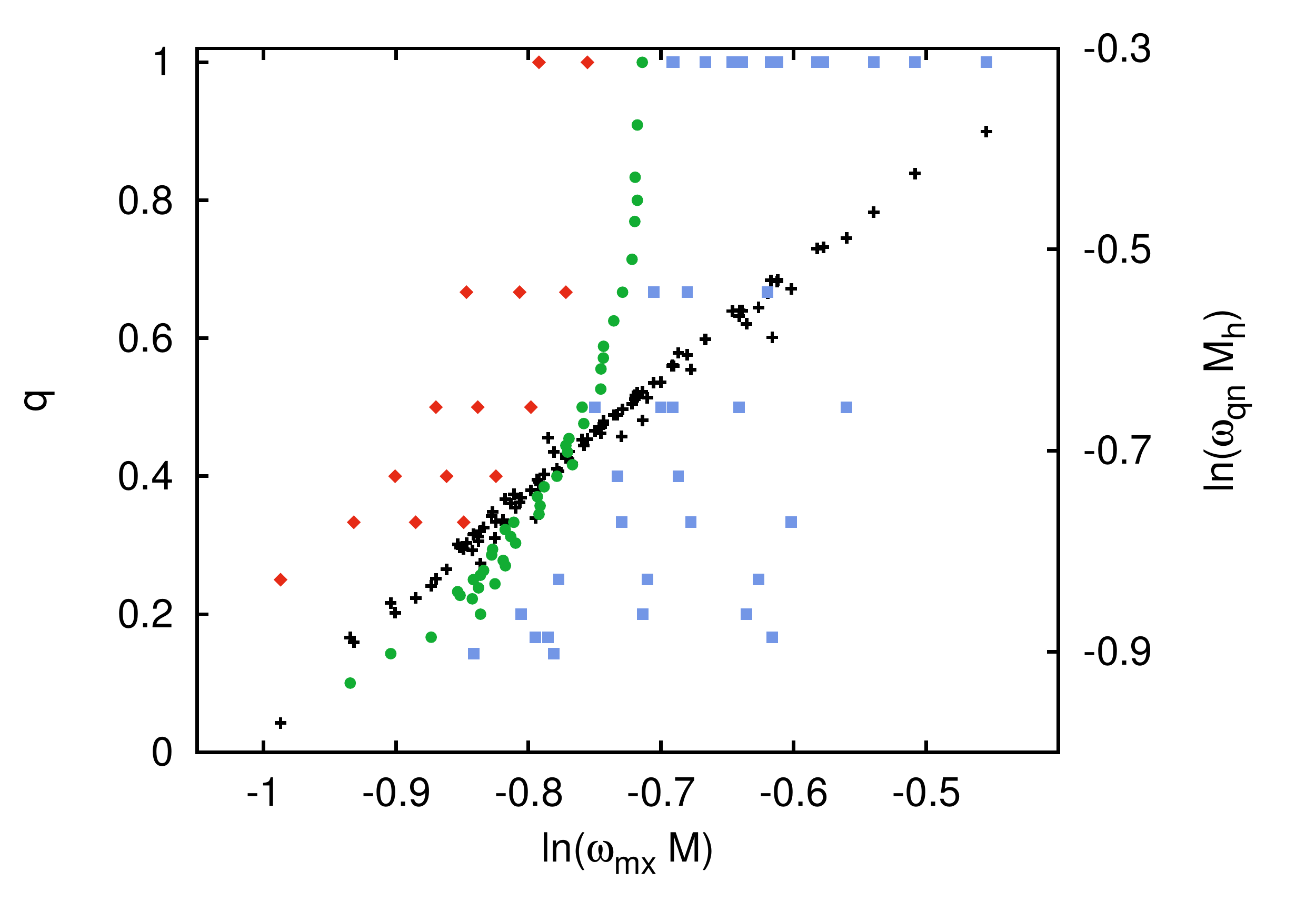}
}
\caption{ Black crosses are the same points as in the top panel of Fig.~\ref{fig:QNMVommax} for the (2,2) mode but with the vertical axis on the right. The points are replotted according to their mass ratio value $q$ (left axis) with a shape and color according to their total spin: non-spinning [green circles], aligned [blue squares], and anti-aligned 
[red diamonds] with the orbital angular momentum. }
\label{fig:NPbreakdown}
\end{figure}

\emph{Connecting $\omega_{\rm mx}$ with \qnm{} Ringing:} 
Fig.~\ref{fig:QNMVommax} shows,  from top to bottom,  the dimensionless quantities $\hat\omega_{\rm qn}  \equiv \omega_{\rm qn}\,M_{\rm h}$, $\hat \tau_{\rm qn} \equiv \tau_{\rm qn}/M_{\rm h}$ and $Q$ as a function of  $\hat \omega_{\rm mx} \equiv \omega_{\rm mx}\,M$ for non-precessing binaries. Each point represents one simulation, with points clustered according to their mode:  blue squares for the (2,2) mode, red triangles for the (3,3) mode, green circles for the (4,4) mode and purple crosses for the (5,5) mode. These are the most dominant  modes of \gw{} emission. 
In order to identify the type of simulation in Fig.~\ref{fig:NPbreakdown}, the same points are replotted according to their mass ratio value $q$ (left axis) with a shape and color according to their total spin: non-spinning as green circles, aligned as blue squares, and anti-aligned with the orbital angular momentum as red diamonds. 
The (2,2) mode data of the top panel of Fig.~\ref{fig:QNMVommax} are also included as black crosses with the vertical axis
on the right.

For each mode, it is evident in Fig.~\ref{fig:QNMVommax} the correlation of $\hat \omega_{\rm mx}$ with  $\hat \omega_{\rm qn}$, $\hat \tau_{\rm qn}$ and $Q$. Moreover, the data seem to imply that 
\begin{eqnarray}
  \ln( \hat \omega_{\rm qn} ) &=& f_1 + f_2 \ln( \hat \omega_{\rm mx} )\label{eq:w}\\
\ln(\hat \tau_{\rm qn}) &=& g_1 + g_2 \ln( \hat \omega_{\rm mx} ) + g_3[ \ln( \hat \omega_{\rm mx} )]^2 \label{eq:t}\\
  \ln( Q ) &=& h_1 + h_2 \ln(\hat  \omega_{\rm mx} ) + h_3 [\ln( \hat \omega_{\rm mx} )]^2\label{eq:q}\,.
\end{eqnarray} 
In Table~\ref{tab:fitting}, we report the values of the fitting parameters for each mode. Notice that $f_2 \simeq 1$, thus 
$
\hat\omega_{\rm qn} \simeq e^{f_1} \hat\omega_{\rm mx}\,.
$

\begin{table}
\begin{center}
\begin{tabular}{lcccc}
Mode   & 22 & 33 & 44 & 55 \\
\hline
$f_1$ & 0.1364 & 0.0926 & 0.1414 & 0.1454 \\
$f_2$ & 1.0985 & 0.9965 & 1.0304 & 1.0164 \\
$g_1$ & 3.5564 & 2.7419 & 2.4850 & 2.3677 \\
$g_2$ & 2.3279 & 1.2249 & 0.7520 & 0.2835 \\
$g_3$ & 1.2044 & 1.0502 & 1.1775 & 0.8340 \\
$h_1$ & 2.8479 & 2.1287 & 1.9369 & 1.8130 \\
$h_2$ & 3.0018 & 2.1433 & 1.7487 & 1.4528 \\
$h_3$ & 0.9130 & 0.9370 & 0.7258 & 0.2841 \\
\end{tabular}
\end{center}
\caption{ Fitting coefficients for Eqs.~(\ref{eq:w}, \ref{eq:t}) and (\ref{eq:q}).}
\label{tab:fitting}
\end{table}

\begin{figure}[tb]
\centering
\vbox{
    \includegraphics[angle=270,width=.68\linewidth]{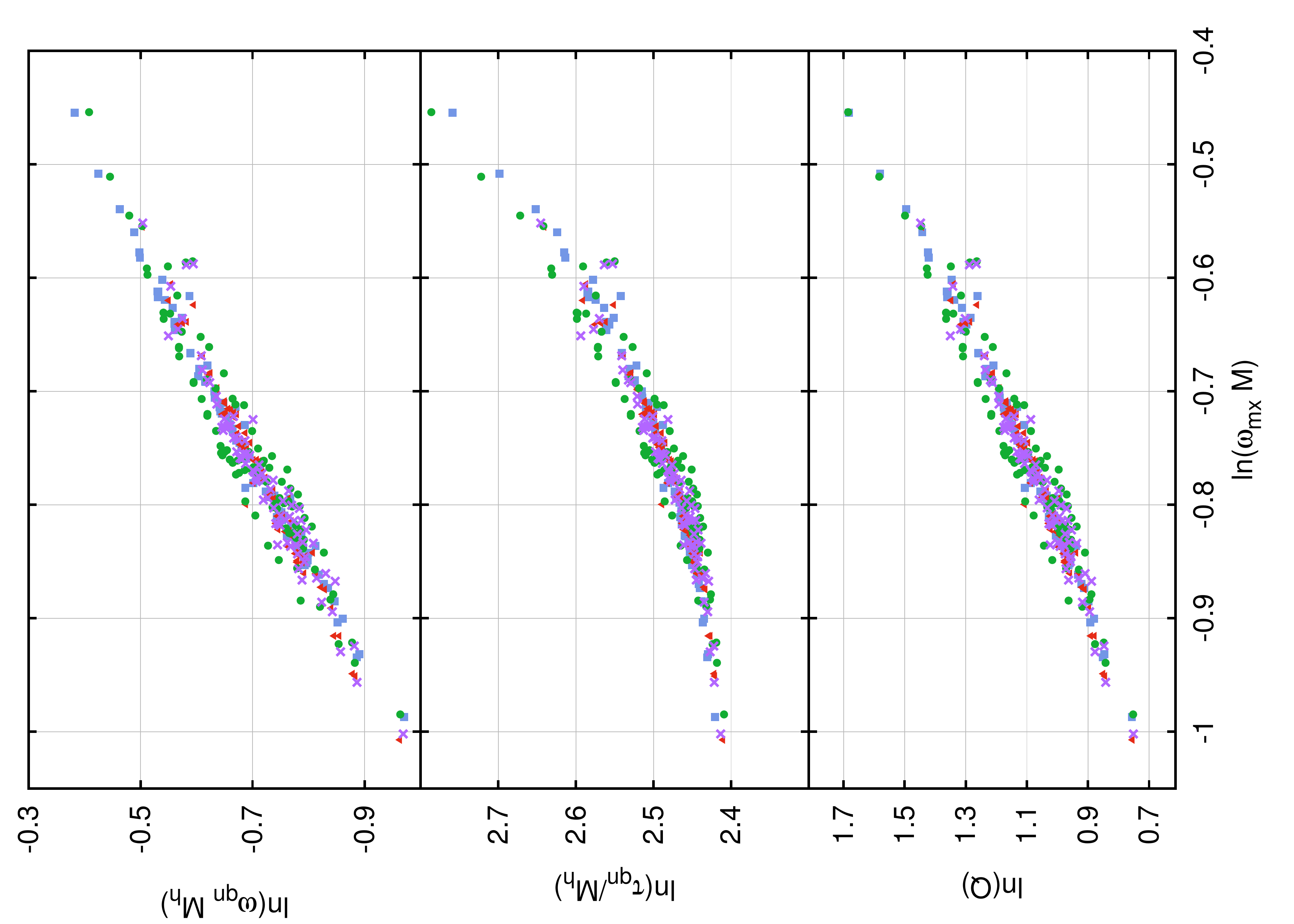}
}
\caption{Same as in Fig.~\ref{fig:QNMVommax} but with the data from the (3,3) [triangles, red], (4,4) [circles, green], and (5,5) [crosses, purple] modes shifted to lie on top of the (2,2) [squares,blue] mode. The values of the shifts are reported in Table~\ref{tab:diffs}.}

\label{fig:shiftQNMVommax}
\end{figure}

Furthermore, there seems to also be a self-similarity among  the modes; that is, with the appropriate shifts, it is possible to cluster all the modes, while still preserving their original correlation characteristics.  This is shown in Fig.~\ref{fig:shiftQNMVommax} where the data from modes (3,3), (4,4) and (5,5) have been shifted to lie on top of the (2,2) mode. The shifts are calculated as the average of the shift for each simulation. For instance, for 
 $\ln(\hat\omega_{\rm qn})$, the shift for the (3,3) mode is obtained from
\begin{equation}
\Delta\ln(\hat\omega_{\rm qn})^{(3,3)} = \langle \ln(\hat\omega_{\rm qn})|^{(3,3)}  - \ln(\hat\omega_{\rm qn}) |^{(2,2)} \rangle\,,
\label{eq:shift}
\end{equation}
where the angle brackets denote average over all the simulations.
The values of the shifts are reported in Table~\ref{tab:diffs}.
Notice from Eqs.~(\ref{eq:w}) and (\ref{eq:shift}) that 
\begin{equation}
\left[\Delta\ln(\hat\omega_{\rm qn}) -  \Delta\ln(\hat\omega_{\rm mx})\right]^{(3,3)} = f_1^{(3,3)}-f_1^{(2,2)}\,,
\label{eq:shift2}
\end{equation}
where we have used that $f_2^{(3,3)} \simeq f_2^{(2,2)}\simeq 1$. The values reported in  Tables~\ref{tab:fitting} and~\ref{tab:diffs} are consistent with Eq.~(\ref{eq:shift2}). Namely,
$\left[\Delta\ln(\hat\omega_{\rm qn}) -  \Delta\ln(\hat\omega_{\rm mx})\right]^{(3,3)} = -0.035$ and $ f_1^{(3,3)}-f_1^{(2,2)} = -0.0414$, with the difference related to the goodness of the fit given by Eq.~(\ref{eq:w}).

\begin{table}
\begin{center}
\begin{tabular}{lcccc}
Mode   & $\Delta\ln (\hat\omega_{\rm mx})$ & $\Delta\ln(\hat\omega_{\rm qn})$ & $\Delta \ln(\hat\tau_{\rm qn})$ & $\Delta\ln(Q)$ \\
\hline
33     & $-0.4299 $ & $-0.4647 $ & $0.0311 $ & $-0.4336 $\\
44     & $-0.6887 $ & $-0.7658 $ & $0.0457 $ & $-0.7200 $\\
55     & $-0.9068 $ & $-0.9952 $ & $0.0577 $ & $-0.9376 $
%
\end{tabular}
\end{center}
\caption{Values of the shifts used in Fig.~\ref{fig:shiftQNMVommax} to bring the
the data from the (3,3), (4,4) and (5,5) modes to lie on top of the (2,2) mode.}
\label{tab:diffs}
\end{table}

\begin{figure}[tb]
\centering
\vbox{
    \includegraphics[angle=270,width=.70\linewidth]{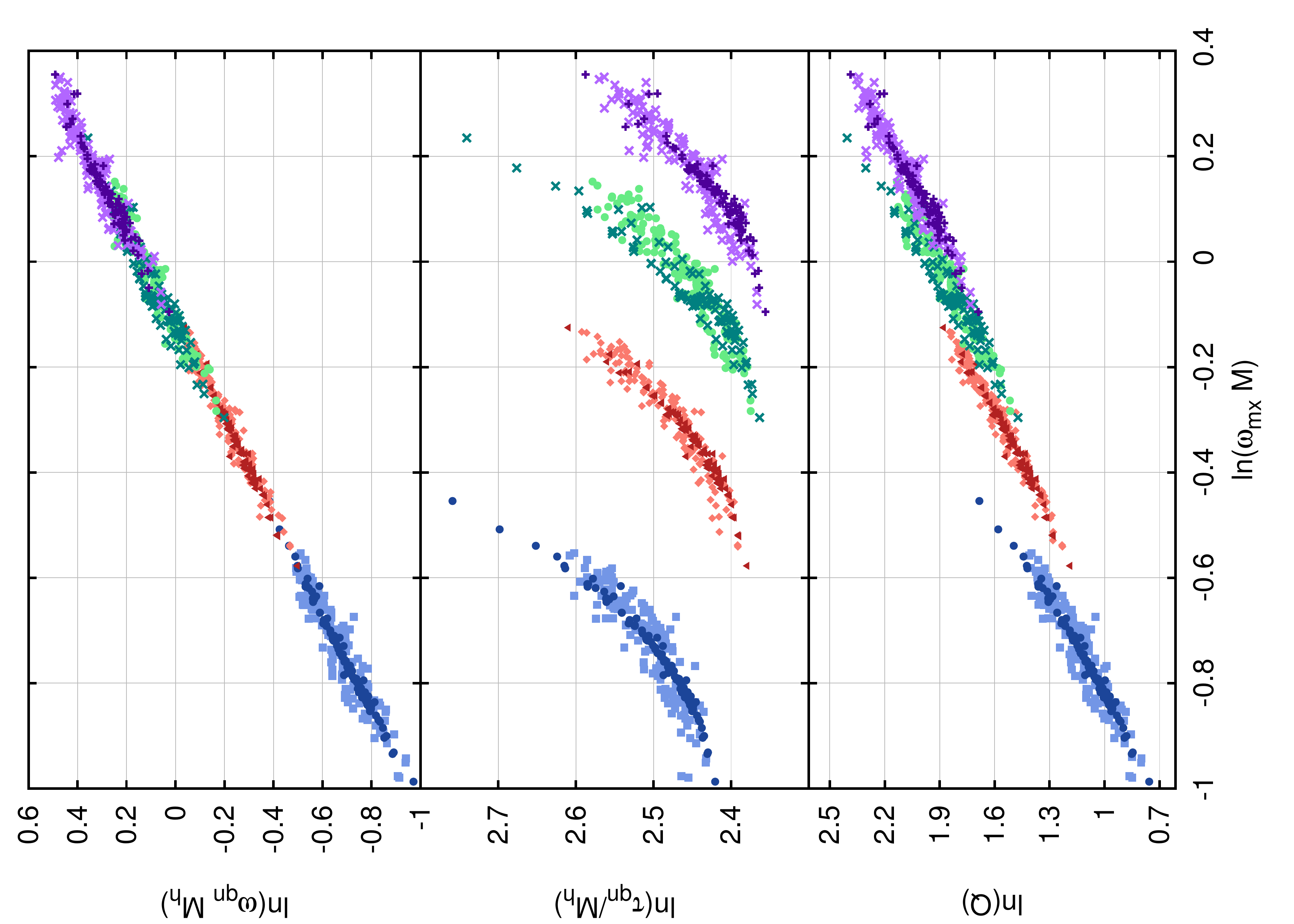}
}
\caption{Same as Fig.~\ref{fig:QNMVommax} but also including precessing binaries. Non-precessing are denoted with dark symbols and precessing with light symbols. Notice increase of
the spread.}
\label{fig:PrecQNMVommax}
\end{figure}

Up until this point, we have only included non-precessing \bbh{s}. The effect of precession is shown in Fig.~\ref{fig:PrecQNMVommax}, where we have included the remaining 172 simulations with precessing binaries. Non-precessing are denoted with dark symbols and precessing with light symbols. It is clear that, although the shapes are preserved, there is a noticeable increase in the spread of the points.

\begin{figure}[tb]
\centering
\vbox{
    \includegraphics[angle=270,width=.68\linewidth]{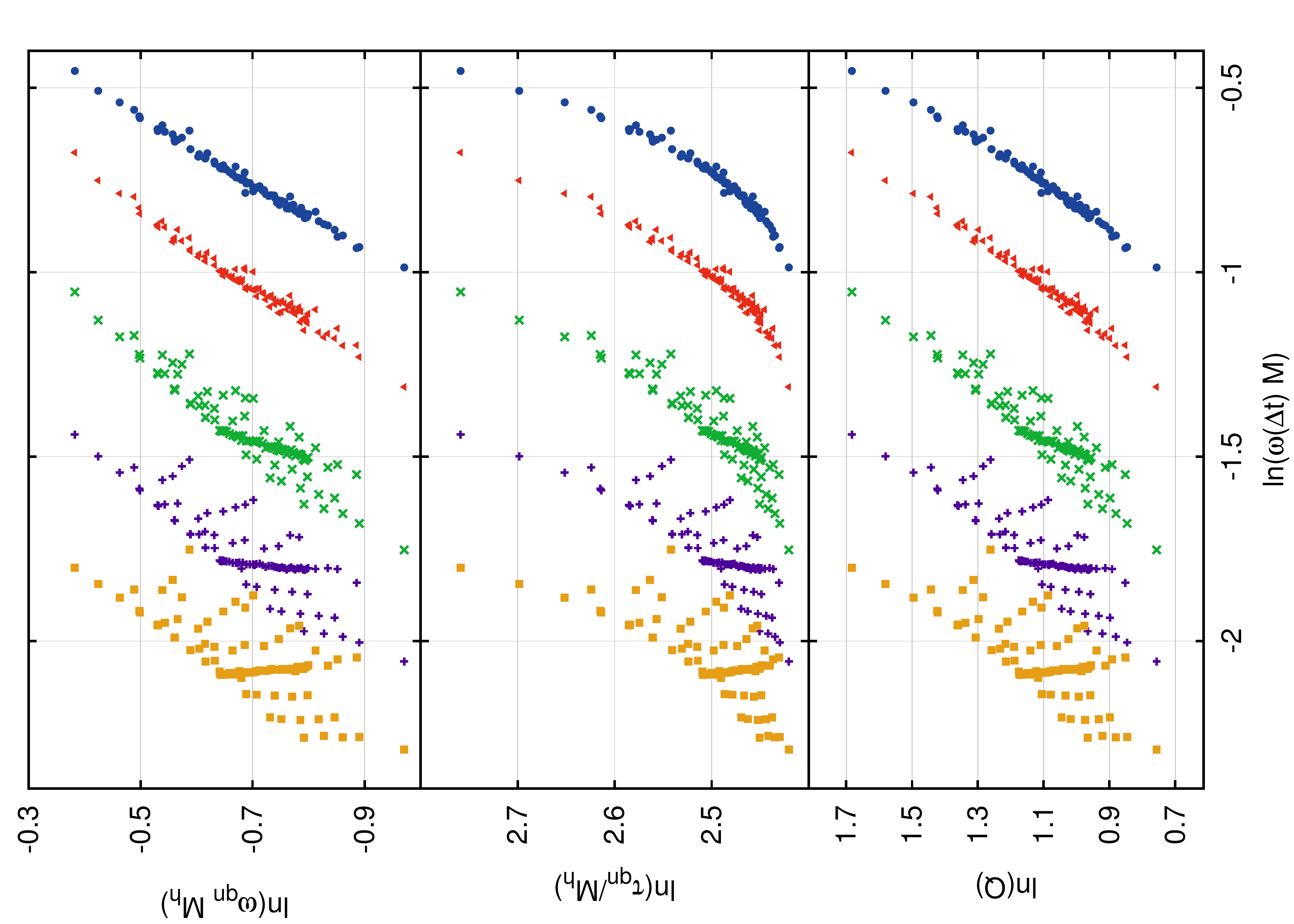}
}
\caption{The quasi-normal mode ringdown frequency
$\hat \omega_{\rm qn}$ (top), the decay time $\hat \tau_{\rm qn}$
(middle), and the quality factor $Q$ (bottom)
versus $\hat \omega(\Delta t)$, where $\hat \omega(\Delta t)$ denotes the frequency 
 of the \gw{} at time $\Delta t$ before $\Psi_4$ peaks for the 
(2,2) mode. From left to right $\Delta t = \{100, 50, 25, 10, 0\}\,M$.
}
\label{fig:QNMVom_of_t}
\end{figure}

Lastly, it is interesting to investigate whether or not $\hat \omega_{\rm mx}$ is special. To do so, we have repeated the analysis, but instead of using $\hat \omega_{\rm mx}$, we use the frequency of the \gw{} at a time $\Delta t$ before peak amplitude. The results for the $(2,2)$  mode are depicted in Fig.~\ref{fig:QNMVom_of_t}. The clusters of points from left to right are for  $\Delta t = \{100, 50, 25, 10, 0\}\,M$, respectively. There is a good indication that  the correlation persists for $\Delta t \le 10\,M$, but very likely not around merger, i.e. $\Delta t \simeq 25\,M$, or earlier. 

\emph{Discussion:} 
Our \bbh{} simulations have unveiled a correlation between the 
frequency $\hat\omega_{\rm mx}$ of the \gw{} around peak amplitude and the \qnm{} ringing of the final \bh{.} The correlation could prove helpful in the construction of templates and data analysis. Furthermore, the correlation of $\omega_{\rm qn}$ and $\tau_{\rm qn}$ with  $\omega_{\rm mx}$  could in principle be used in both directions. Namely, if one is able to estimate $\omega_{\rm mx}$ in the \gw{} via for instance excess-power, our correlation could tell us about the final \bh{.} But one could also envision, given the mass and spin of the final \bh{} of interest, using the correlation to constrain the frequency of the \gw{} soon after merger, when the power of emission is the strongest.  Specifically, for \bbh{} with large masses, e.g.  $50\,M_\odot \le M  \le 500\,M_\odot$, only the last few cycles, merger and ring-down lie within the sweet spot of the detector. The search basically reduces to that of a perturbed intermediate mass \bh{}~\cite{2014PhRvD..89j2006A} for which sine-gaussians or chirplets are useful representations of the signal~\cite{2010CQGra..27s4017C}. The correlations in our work could be used to fix the characteristic frequency of the chirplet, i.e.  $\omega_{\rm mx}$, 
given the range of masses and spins of the final \bh{.} 

Further, recent work has shown a connection between the ringdown properties and the initial \bh{} parameters\cite{Kamaretsos:2012bs,London:2014cma,Meidam:2014jpa}.
Our current work would extend this, allowing both the final state and the initial state of the system to be identified by
just the burst region of the waveform.

The correlation between $\hat\omega_{\rm mx}$ and the \qnm{}  could also aid in the tuning of phenomenological models to \bbh{} mergers\cite{Buonanno:2009qa,Ajith:2009bn,Hinder:2013oqa}.   The relationship given in Eqs.~(\ref{eq:w}), (\ref{eq:t}) and (\ref{eq:q}), provides a method for  adding ringdown to the waveform using just the peak frequency  of the waveform. Perhaps most importantly, the method applies to all the modes excited during merger.

\ack
PL and DS supported by NSF grants 0955825, 1205864, 1212433, 1333360, and JH by NSF grants 1305730 and 0969855. Computations at XSEDE PHY120016 and the Cygnus cluster at Georgia Tech.  Authors thank Lionel London for useful discussions.

\section*{References}
\bibliography{references}

\providecommand{\newblock}{}
\begin{thebibliography}{10}
\expandafter\ifx\csname url\endcsname\relax
  \def\url#1{{\tt #1}}\fi
\expandafter\ifx\csname urlprefix\endcsname\relax\def\urlprefix{URL }\fi
\providecommand{\eprint}[2][]{\url{#2}}

\bibitem{0264-9381-27-8-084006}
Harry G~M and the LIGO Scientific~Collaboration 2010 {\em CQG\/} {\bf 27}
  084006

\bibitem{0264-9381-28-11-114002}
Accadia T {\em et~al.\/} 2011 {\em CQG\/} {\bf 28} 114002

\bibitem{PhysRevD.88.043007}
Aso Y {\em et~al.\/} (The KAGRA Collaboration) 2013 {\em Phys. Rev. D\/} {\bf
  88}(4) 043007

\bibitem{4690}
Echeverria F 1989 {\em Phys. Rev. D\/} {\bf 40} 3194--3203

\bibitem{Haas:2012bk}
Haas R, Shcherbakov R~V, Bode T and Laguna P 2012 {\em Astrophys.J.\/} {\bf
  749} 117

\bibitem{Healy:2011ef}
Healy J, Bode T, Haas R, Pazos E, Laguna P {\em et~al.\/} 2011

\bibitem{Bode:2011xz}
Bode T, Laguna P and Matzner R 2011 {\em Phys.Rev.\/} {\bf D84} 064044

\bibitem{Bode:2011tq}
Bode T, Bogdanovic T, Haas R, Healy J, Laguna P {\em et~al.\/} 2012 {\em
  Astrophys.J.\/} {\bf 744} 45

\bibitem{Bode:2009mt}
Bode T, Haas R, Bogdanovic T, Laguna P and Shoemaker D 2010 {\em Astrophys.
  J.\/} {\bf 715}

\bibitem{Healy:2009zm}
Healy J, Levin J and Shoemaker D 2009 {\em Phys. Rev. Lett.\/} {\bf 103} 131101

\bibitem{et-web}
Einstein Toolkit home page:{\tt http://www.einsteintoolkit.org}

\bibitem{cactus-web}
Cactus Computational Toolkit home page:{\tt http://www.cactuscode.org}

\bibitem{Schnetter-etal-03b}
Schnetter E, Hawley S~H and Hawke I 2004 {\em Class. Quant. Grav.\/} {\bf 21}
  1465--1488

\bibitem{Husa:2004ip}
Husa S, Hinder I and Lechner C 2006 {\em Computer Physics Communications\/}
  {\bf 174} 983--1004

\bibitem{2014PhRvD..89j2006A}
{Aasi} J {\em et~al.\/} 2014 {\em PRD\/} {\bf 89} 102006

\bibitem{2010CQGra..27s4017C}
{Chassande Mottin} {\'E}, {Miele} M, {Mohapatra} S and {Cadonati} L 2010 {\em
  CQG\/} {\bf 27} 194017

\bibitem{Kamaretsos:2012bs}
Kamaretsos I, Hannam M and Sathyaprakash B 2012 {\em Phys.Rev.Lett.\/} {\bf
  109} 141102

\bibitem{London:2014cma}
London L, Healy J and Shoemaker D 2014  (\textit{Preprint} \eprint{1404.3197})

\bibitem{Meidam:2014jpa}
Meidam J, Agathos M, Van Den~Broeck C, Veitch J and Sathyaprakash B 2014
  (\textit{Preprint} \eprint{1406.3201})

\bibitem{Buonanno:2009qa}
Buonanno A, Pan Y, Pfeiffer H~P, Scheel M~A, Buchman L~T {\em et~al.\/} 2009
  {\em PRD\/} {\bf 79} 124028

\bibitem{Ajith:2009bn}
Ajith P, Hannam M, Husa S, Chen Y, Bruegmann B {\em et~al.\/} 2011 {\em PRL\/}
  {\bf 106} 241101

\bibitem{Hinder:2013oqa}
Hinder I, Buonanno A, Boyle M, Etienne Z~B, Healy J {\em et~al.\/} 2014 {\em
  CQG\/} {\bf 31} 025012

\end{thebibliography}

\end{document}